\documentclass[a4paper,11pt]{article}
\pdfoutput=1
\usepackage{jcappub}
\usepackage[T1]{fontenc}

\def\be{\begin{equation}}
\def\ee{\end{equation}}
\newcommand{\beq}{\begin{eqnarray}}
\newcommand{\eeq}{\end{eqnarray}}

\title{\boldmath Gravitational wave echoes from black hole area quantization}

\author[a,b]{Vitor Cardoso,}
\author[c]{Valentino F.\ Foit,}
\author[c]{Matthew Kleban}

\affiliation[a]{Centro de Astrof\'{\i}sica e Gravita\c c\~ao - CENTRA, Departamento de F\'{\i}sica
			\\Instituto Superior T\'ecnico - IST, Universidade de Lisboa, Lisboa, Portugal}
\affiliation[b]{Theoretical Physics Department, 
CERN 1 Esplanade des Particules, Geneva 23, CH-1211, Switzerland}
\affiliation[c]{Center for Cosmology and Particle Physics\\
			New York University, New York, USA}

\emailAdd{vitor.cardoso@ist.utl.pt}
\emailAdd{foit@nyu.edu}
\emailAdd{kleban@nyu.edu}

\abstract{Gravitational-wave astronomy has the potential to substantially advance our knowledge of the cosmos, from the most powerful astrophysical engines to the initial stages of our universe. Gravitational waves also carry information about the nature of black holes. Here we investigate the potential of gravitational-wave detectors to test a proposal by Bekenstein and Mukhanov that the area of black hole horizons  is quantized in units of the Planck area. Our results indicate that this quantization could have a potentially observable effect on the {\it classical} gravitational wave signals received by 
 detectors. In particular, we find distorted gravitational-wave ``echoes'' in the post-merger waveform describing the inspiral and merger of two black holes. These echoes have a specific frequency content that is  characteristic of black hole horizon area quantization.   }

\begin{document} 
\maketitle
\flushbottom

\section{Introduction}
\subsection{Quantization of black hole area}
The quantum mechanics of black hole (BH) event horizons has a number of surprising features, most notably that BHs radiate thermally at the Hawking temperature and that the entropy of a BH is proportional to its horizon area. For astrophysical BHs neither of these effects is likely to be observable. However, a number of authors have proposed that the quantum modifications go well beyond these subtle effects, so that the physics even of large BHs can be dramatically different from the classical theory.  
One such proposal is that of ``firewalls'', where the spacetime just behind the horizon is singular due to quantum effects that induce a very large energy density there~\cite{Almheiri:2012rt}. If true, this would have dramatic effects on an observer falling through the horizon. Modifications of the region just outside the horizon could produce signals visible to external observers, include characteristic electromagnetic bursts when these objects collide~\cite{Abramowicz:2016lja,Kaplan:2018dqx}, or late-time ``echoes'' in gravitational-wave signals~\cite{Barausse:2014tra,Cardoso:2016rao,Cardoso:2016oxy,Cardoso:2017cqb}.

A  perspective pioneered by Bekenstein and Mukhanov~\cite{Bekenstein:1974jk,Bekenstein:1995ju} is that the  horizon area $A$ of BHs is quantized in units of the Planck area $ l_{P}^{2}$:
\be
	A = \alpha l_{P}^{2} N = \alpha \frac{\hbar G}{c^2} N. \label{quant}
\ee
Here $N$ is an integer and $\alpha$  an $\mathcal{O}(1)$ dimensionless coefficient.  

One might expect that quantization of horizon area in such tiny units would have no observable implications for realistic BHs. However, taken at face value Eq.~\eqref{quant} implies that the spectrum of emission or absorption of radiation by quantum BHs occurs in a series of evenly-spaced lines. For a Schwarzschild BH of mass $M$, $A = 4 \pi r_s^2 = 4 \pi (2 G M/c^2)^{2}$, so 
$$
	\Delta A =  \alpha \frac{\hbar G}{c^3}  \Delta N =  32 \pi  \frac{G^{2}}{c^4} M \Delta M
$$ 
and therefore the frequency absorbed or emitted by a BH is quantized by
\be
	\omega_{n} = \left| \frac{\Delta M}{\hbar} \right| = \frac{n \alpha}{32 \pi} \frac{c}{M G} = \frac{n \alpha}{16 \pi} {c \over r_s} \,.\label{quant2}
\ee
Here $n = \left| \Delta N \right|$ is the change in the area quantum. 

Various values have been suggested for the proportionality constant $\alpha$ in Eq.~\eqref{quant}. The original
argument requires that the number of states $e^{S}$ be an integer, leading to $\alpha = 4 \ln q$, where $q$ is an integer~\cite{Bekenstein:1995ju}. Different arguments lead to different integers, for example $q=2, 3$~\cite{Mukhanov:1986me,Hod:1998vk,Dreyer:2002vy}. A matching to highly damped quasinormal modes (QNMs)~\cite{Berti:2009kk} leads to $\alpha = 8 \pi$~\cite{Maggiore:2007nq,Vagenas:2008yi, Medved:2008iq}. A ``holographic shell model'' for BHs suggests $\alpha = 8 \ln 2$~\cite{Davidson:2011eu}. In summary, $1 < \alpha < 30$ seems a reasonable expectation.

\subsection{Imprints on classical observables}
If the Bekenstein-Mukhanov proposal is correct, BHs behave more like simple atoms than large, opaque objects~\cite{Bekenstein:1997bt}.  In particular they are optically thin and can only absorb (or emit) radiation with wavelengths close to integer multiples of the fundamental mode \eqref{quant2} \cite{Mukhanov:1986me}.
However, the proposal does not specify what microphysics is responsible for the quantization, nor does it give a detailed prediction for the dynamics in all possible scenarios.  As such, there are several possibilities for precisely how it modifies  BH physics.

There is an effective potential energy that describes the motion of test particles near classical black holes. This potential has a maximum at the ``photon sphere'', which for Schwarzschild BHs is at radius $r =3GM/c^2 \equiv (3/2)r_s$. This barrier partially screens the near-horizon region from outside observers~\cite{Cardoso:2016rao,Cardoso:2017cqb,Cardoso:2017njb}.    One possible scenario is that the quantum modification and  frequency quantization \eqref{quant2} modifies the physics in a large region out to and including the barrier at the photon sphere.  In Ref.~\cite{Foit:2016uxn}, two of us showed that LIGO could test this scenario.  Radiation emitted to infinity during the initial ringdown phase following a merger that produces a BH (or other perturbation to an existing BH) would be affected by the quantization in a fairly simple way. Because of the uncertainty  in the value of the unknown parameter $\alpha$, the observation of the ringdown from a single merger event might not suffice to test the Bekenstein-Mukhanov proposal, but would determine $\alpha$. A second ringdown observation with a differing value for the spin parameter of the final BH would then suffice to rule  out the proposal (or provide strong evidence for it)~\cite{Foit:2016uxn}.  This is because the fundamental frequency (the analog of \eqref{quant2} for a spinning BH) depends on the spin in a different way than the the spin dependence of the primary ringdown frequency that would be emitted according to classical physics.

A second possibility is that the quantum modification affects only the region just outside  the horizon.  To be definite, suppose the modification to classical physics is important only at radii $r < r_\epsilon={2 G M \over c^2}(1+\epsilon) = r_s ( 1 + \epsilon)$, with $\epsilon \ll 1$. This possibility was left to future work in Ref.\ \cite{Foit:2016uxn}, and will be our sole focus here. Radiation  that falls towards the horizon will be reflected near $r = r_\epsilon$ if it has  the ``wrong'' frequency, while radiation with the frequencies \eqref{quant2} will be absorbed. BHs would no longer be perfect absorbers.  This property could show up both in the inspiralling stage of BH binaries and during the last stages of BH relaxation after a merger or other large perturbation. 

\subsubsection{Gravitational-wave echoes}
In the classical theory, the first part of the ringdown signal following a merger is controlled by the characteristics of the aforementioned photon sphere potential barrier that for a non-spinning BH has radius $r = (3/2)r_s$. In the scenario we are considering -- where the quantum modifications are confined to a region very close to the horizon at $r = r_{s}$ -- this initial signal will be unaffected by the area quantization. However, the partially reflecting properties
of the horizon do give rise to new phenomena: some radiation that would have fallen into the horizon is now reflected back.  It then interacts with the potential barrier at the photon sphere and is partially transmitted and partially reflected back towards the horizon, generating a series of so-called gravitational-wave echoes~\cite{Cardoso:2016rao,Cardoso:2016oxy}. These pulses will be separated in time by a delay due to the time it takes the signal to transit from the horizon to the photon sphere. This delay time scales logarithmically with $\epsilon$, and for a solar mass BH with $\epsilon =l_P/r_s$ it is a few milliseconds, much longer than the ringdown timescale of roughly 10 $\mu$s.  This makes it easy to separate the echoes from each other and from the initial ringdown phase.

Echoes from exotic compact objects have been investigated quite extensively theoretically in the last few years~\cite{Cardoso:2016rao,Cardoso:2016oxy,Mark:2017dnq,Correia:2018apm,Bueno:2017hyj,Cardoso:2017cqb,Cardoso:2017njb}, and searches for the presence of echoes in gravitational-wave data are being implemented~\cite{Abedi:2016hgu,Abedi:2017isz,Conklin:2017lwb,Westerweck:2017hus,Tsang:2018uie,Nielsen:2018lkf,Lo:2018sep}. However, no compelling theoretical motivation was provided for these earlier studies. Rather, researchers posited some specific modification (for instance, that there is a perfectly reflecting boundary condition a Planck length outside the Schwarzschild radius) and studied the resulting spectrum of echoes. Here for the first time we study the echoes  produced by a long-standing and relatively well-motivated proposal  \cite{Bekenstein:1974jk,Bekenstein:1995ju} for a modification of the horizon physics, namely  area quantization \eqref{quant2}.   As we will see, the echoes are filtered by the frequency-dependent absorption in a way that, if observed, could confirm the hypothesis of BH horizon area quantization.

\subsubsection{Inspiral} Modifying horizon physics could change the inspiral phase of a merger as well. Consider two BHs inspiralling around each other. Most of the emitted gravitational radiation moves outwards, but a fraction interacts with the BHs. As a consequence of energy conservation and the quasi-adiabatic evolution of the geodesic parameters, the total energy loss $\dot{E}=\dot{E}^\infty+\dot{E}^{\rm horizon}$ determines how quickly the inspiral proceeds \cite{Poisson:1994yf,Hughes:2001jr}.  One  channel for the orbit to lose energy is through the horizon. If the horizon is replaced by a reflective surface,  the rate at which the inspiral proceeds is decreased since that channel is closed. For spinning objects this effect appears at ($2.5 \times \log {\rm velocity}$) post-Newtonian order, and can be used to  constrain modifications of the near-horizon physics \cite{Maselli:2017cmm}. The calculations of \cite{Maselli:2017cmm} were kept general and apply in principle to  near-horizon area quantization.  An interesting possibility, left to future work, is that the inspiral  may excite  resonances of the near-horizon region of the black hole.

From here forward we use natural units $c=G=\hbar=1$.

\section{The quantum filter}
\subsection{Framework}
%
\begin{figure}[t]
\centering
\includegraphics[]{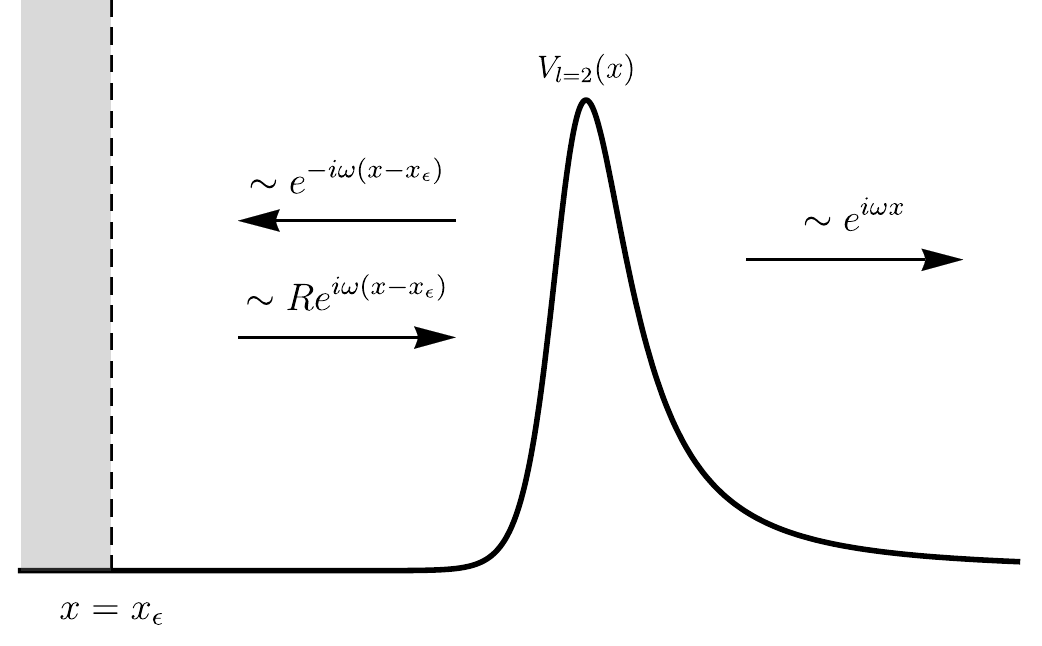}
\caption{The effective gravitational potential \eqref{veff} with $s=l=2$ is shown in terms of the tortoise coordinate $x$. At $x=x_\epsilon$ there is a reflective surface with reflection coefficient $R(\omega)$. The boundary conditions for purely outgoing waves at infinity are shown.}
\label{fig:veff4d_alt}
\end{figure}
We focus entirely on the regime which  describes the late-time dynamics of post-merger binaries.
In particular, the geometry is taken to depart only slightly from a BH vacuum solution of the field equations.
For simplicity we consider only non-spinning BHs, but our results can straightforwardly be extended to Kerr geometries. Then, the dynamics can be linearized and expanded in
spin-$s$ spherical harmonics, with $s=0,1,2$ for scalar field, electromagnetic or gravitational fluctuations~\cite{Berti:2009kk}. It is possible to show that all massless fields are governed by the master equation
\be
	\left[ \partial_t^2 - \partial_x^2 +  V_l(r) \right] \psi(x,t) = {\cal S}(x,t)\,,\label{waveeq}
\ee
where the effective potential is given by
\be
	V_l(r) = \frac{l(l+1)}{r^2} + f\frac{(1-s^2)2M}{r^3}~.
	\label{veff}
\ee
Here, $f=1-2M/r$ and the tortoise coordinate $x$ is related to the Schwarzschild coordinate $r$ as
\be
	x = \int^r \frac{d r'}{f(r')} = r + 2M \log\left(\frac{r}{2 M}-1 \right)\,.
\ee
The source term ${\cal S}(x,t)$ is non-zero when there is a ``charge'' (such as an electromagnetic charge, or a mass) in the BH exterior.
The source term will be set to zero when we study scattering of massless fields in Section \ref{sec:scatters}, but is non-vanishing and describes ``small'' BHs or compact stars falling into a large BH in Section \ref{sec:inspiral}.

By causality the field obeys outgoing Sommerfeld conditions at large $x$, $\partial_t \psi+\partial_x\psi=0$.
For stationary backgrounds such as the ones we consider, the field can be Fourier analyzed which yields the Schr{\"o}dinger-like equation
\begin{align}
	\left[\partial_x^2 + \omega^2 - V_l(r)\right] \tilde{\psi}(x,\omega) = -\tilde{{\cal S}}(x,\omega).
\end{align}
The Sommerfeld conditions at infinity are equivalent to
\begin{align}
	\tilde{\psi} (x) \propto e^{i \omega x}\ \quad \text{as} \quad x\to\infty.
\end{align}

As discussed above, we assume that the BH physics is modified near the horizon at $r < r_\epsilon \equiv 2 M(1 + \epsilon)$, for some $\epsilon \ll 1$. For our purposes we will treat the surface  $r=r_\epsilon$ as a boundary where the field is forced to satisfy the nontrivial boundary condition \cite{Mark:2017dnq}
\begin{align}
	\tilde{\psi}(x_\epsilon) \propto e^{-i \omega (x - x_\epsilon)} + R(\omega) e^{i \omega (x - x_\epsilon)}.
\end{align}
Here, $R$ is the reflection coefficient of the surface, see Fig.\ \ref{fig:veff4d_alt}. The $R=0$ limit reproduces the standard GR calculation. In the scenario we are considering the reflection coefficient $R(\omega)$ is close to unity, except around the special values given by \eqref{quant2}. 
The exact width of these absorption lines was left unspecified, but was assumed to be much less than the energy gap between the lines.  For instance, taking the ratio of  the width to the gap to be $\sim 1/30$ reproduces the luminosity of Hawking radiation to a single species of scalar particle \cite{Mukhanov:1986me}.

To summarize, BH horizon area quantization effectively introduces a surface at  a radius $2M(1+\epsilon)$ which absorbs certain wavelengths and reflects the rest.
For standard inspiral and merger processes, just after the merger there will be an initial ringdown signal which is determined by the properties of the photon sphere \cite{Cardoso:2016rao,Cardoso:2016oxy,Cardoso:2017njb}. This signal is identical to that of classical GR, independent of the modification at $r_\epsilon$. After a time $\sim 4M\log\epsilon^{-1}$, waves caused by reflections at $r_\epsilon$ arrive at the photon sphere. Some of these will be transmitted, resulting in the first echo, and some will be reflected back towards the horizon. The end result is a series of echoes of the original burst~\cite{Cardoso:2016rao,Cardoso:2016oxy,Cardoso:2017njb,Mark:2017dnq,Correia:2018apm}. The spectrum of the $k$th echo will have a characteristic shape given that it was ``filtered'' $k$ times by the frequency-dependent reflection near the horizon. We now proceed to analyze the shape of this filtered signal.

\begin{figure}[th]
\begin{tabular}{cc}
\includegraphics[width=0.45\textwidth]{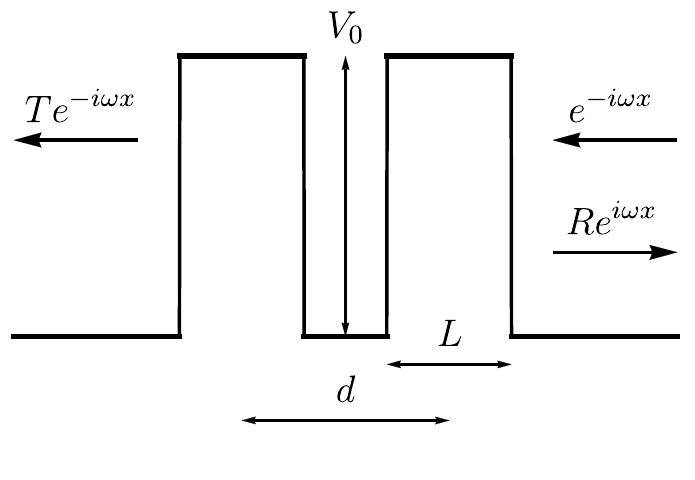}&\includegraphics[width=0.45\textwidth]{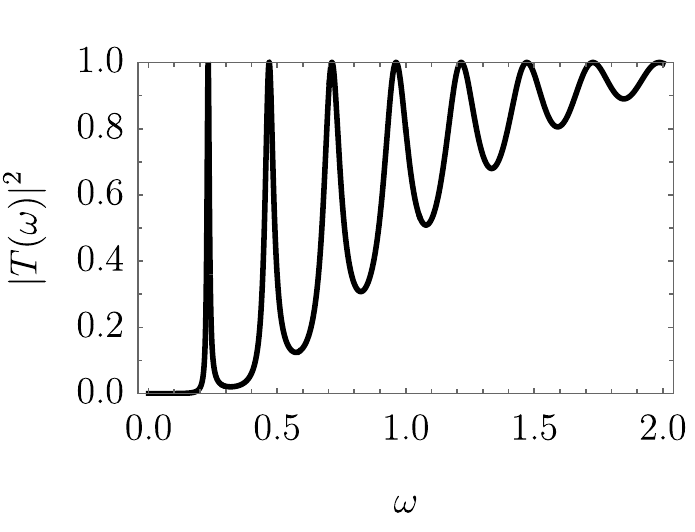}
\end{tabular}
\caption{Left: A double square barrier potential with incoming radiation from infinity. $R$ and $T$ are the (complex) reflection and transmission coefficients, respectively. Right: The transmission probability $|T|^2$ as a function of frequency with parameters $d=12, L=1, V_0=1$.}
\label{fig:doublesquarebarrierpotential}
\end{figure}

\subsection{The double barrier filter\label{subsec:doublebarrier}}

One way to prescribe appropriate boundary conditions for Eq.~\eqref{waveeq} near the horizon is to add an additional potential energy term that has the desired reflection coefficient. The scattering problem can then be solved numerically.  

According to \eqref{quant2}, absorption of radiation near the horizon should occur only in a series of fairly sharp, roughly evenly-spaced lines.
A double square barrier potential with an appropriate set of parameters mimics this behavior.
For nearly evenly spaced frequencies, the double barrier admits perfect transmission, while other frequencies are mostly reflected, Fig.~\ref{fig:doublesquarebarrierpotential}.

The complex reflection and transmission coefficients $R(\omega)$ and $T(\omega)$ are uniquely determined by imposing continuity of the wave and its first spatial derivatives across the potential. The results are known exactly, but the  expressions are too lengthy to reproduce here. The reflection and transmission probability for a wave of frequency $\omega$ are given by $|R(\omega)|^2$ and $|T(\omega)|^2$, respectively.

The distance $d$ between the square barriers, their widths $L$, and their height $V_0$
control the spacing and the widths of the lines in the transmission coefficient.
The functional dependence can be obtained  perturbatively.
For $\omega_n L \ll 1$ the maxima in the transmission probability occur with approximately equidistant real parts
\begin{align}\label{spacing}
	\operatorname{Re} \omega_n \simeq \frac{n \pi}{d-L+\frac{2}{\sqrt{V_0}}\coth(L\sqrt{V_0})},
\end{align}
while the line widths scale like
\begin{align}
	\operatorname{Im} \omega_n \sim \frac{n^4}{V_0^4 L^4 d^6},
\end{align}
which can be substantially smaller than the gaps (as expected for the BH case).

\subsubsection{The choice of  parameters}

Quantization of the horizon area is determined up to a free parameter $\alpha$, which determines the spacing of the lines in the spectrum \eqref{quant2}.  As discussed above, the proposals in the literature for $\alpha$ range roughly from 1 to 30. In the following, we choose $\alpha = 4\pi$, so that $\omega_n = \frac{n}{8 M}$.
We then chose the parameters of the double well potential such that \eqref{spacing}   reproduces these values. A plot of the transmission probability which mimics the area quantization is shown in Fig.\ \ref{fig:doublesquarebarrierpotential}.  

From this point forward we use units with $2 M = 1$.

\subsection{Bandpass filter method\label{subsec:bandpass}}

Naturally, we do not expect the specific characteristics of the double square barrier to correspond precisely to the physics of a BH with quantized area under the Bekenstein-Mukhanov proposal.  Therefore we  also describe a more general method:  a simple approximation that allows us to estimate the effect of any given reflection coefficient $R$ at the modified horizon on the echoes.  Roughly speaking, the Fourier transform of the $k$th echo is multiplied by $R^k$ due to having been reflected $k$ times off the horizon.  This procedure has been described and utilized in the study of gravitational wave echoes, for example in \cite{Mark:2017dnq}.

Using the analytic reflection coefficient $R$ for the double barrier, we  can test this approximation against the numerical approach.  The two  are in good agreement when the problem admits separation of scales ($\epsilon \ll 1$), as we show  in Section~\ref{filtering_compare_gaussian}.  If echoes are  detected in data, this approximation allows the signal to be ``inverted'', producing (an approximation to) the reflection coefficient at the horizon, which could then be compared to the Bekenstein-Mukhanov proposal.

\section{Echoes from the scattering of wavepackets\label{sec:scatters}}

\subsection{The boundary conditions}

As discussed above, at large distances we consider the BH to be isolated, so only outgoing waves are allowed and Sommerfeld boundary conditions will be used at infinity.
We solve the wave equation~\eqref{waveeq} for quadrupolar gravitational fluctuations $s=l=2$ (usually the dominant components of gravitational radiation), subjected to two different boundary conditions at the ``surface'' $r_\epsilon$:

\noindent {\bf D}: We impose Dirichlet, reflective boundary conditions at $x=x_\epsilon = x(r_\epsilon)$.
	
\noindent {\bf B}: As described in Section~\ref{subsec:doublebarrier}, we add a double barrier at $x_\epsilon$ to filter out the relevant frequencies.  In particular, the total effective potential is taken to be
\be
	V(x) = V_{l=2}(x) + V_\text{B}(x)\,,
\ee
where $V_{l}$ is given in Eq.~\eqref{veff} and
\begin{align}
	V_\text{B}(x) = V_0 \left[\operatorname{H}\left(\left|x-x_\epsilon + \frac{d-L}{2}\right| - \frac{d-L}{2}\right) - \operatorname{H}\left(\left|x-x_\epsilon+ \frac{d-L}{2}\right| - \frac{d+L}{2}\right)
	\right]
\end{align}
is a double barrier potential (see Fig.\ \ref{fig:doublesquarebarrierpotential}) whose right barrier ranges from $x_\epsilon$ to $x_\epsilon+L$, and $\operatorname{H}$ is the Heaviside function. This  placement is chosen for convenience, so that there is no time-delay relative to the Dirichlet boundary conditions at $x_\epsilon$. 

The solutions of the time evolution are denoted according to their boundary conditions as $\psi^{D,B}(x,t)$.
For the time evolution we chose units such that $2M=1$ and the parameters
\begin{align}\label{params}
	d=12,~ L= 1,~ V_0 = 1,~ x_\epsilon = -70.
\end{align}
The initial data is described by
\be \label{scatter_params}
	\psi(x,0) = \frac{1}{3} e^{\frac{(x-x_c)^2}{\sigma^2}}, \quad \left. \partial_t \psi(x,t) \right|_{t = 0} = 0 \\
\ee
with $\sigma = 3, x_c = 220$. We probe the wave response for an observer at $x=150$ after the initial Gaussian packet has scattered off the BH.

\subsection{The scattered pulse}

\begin{figure}[t]
\centering
\includegraphics[]{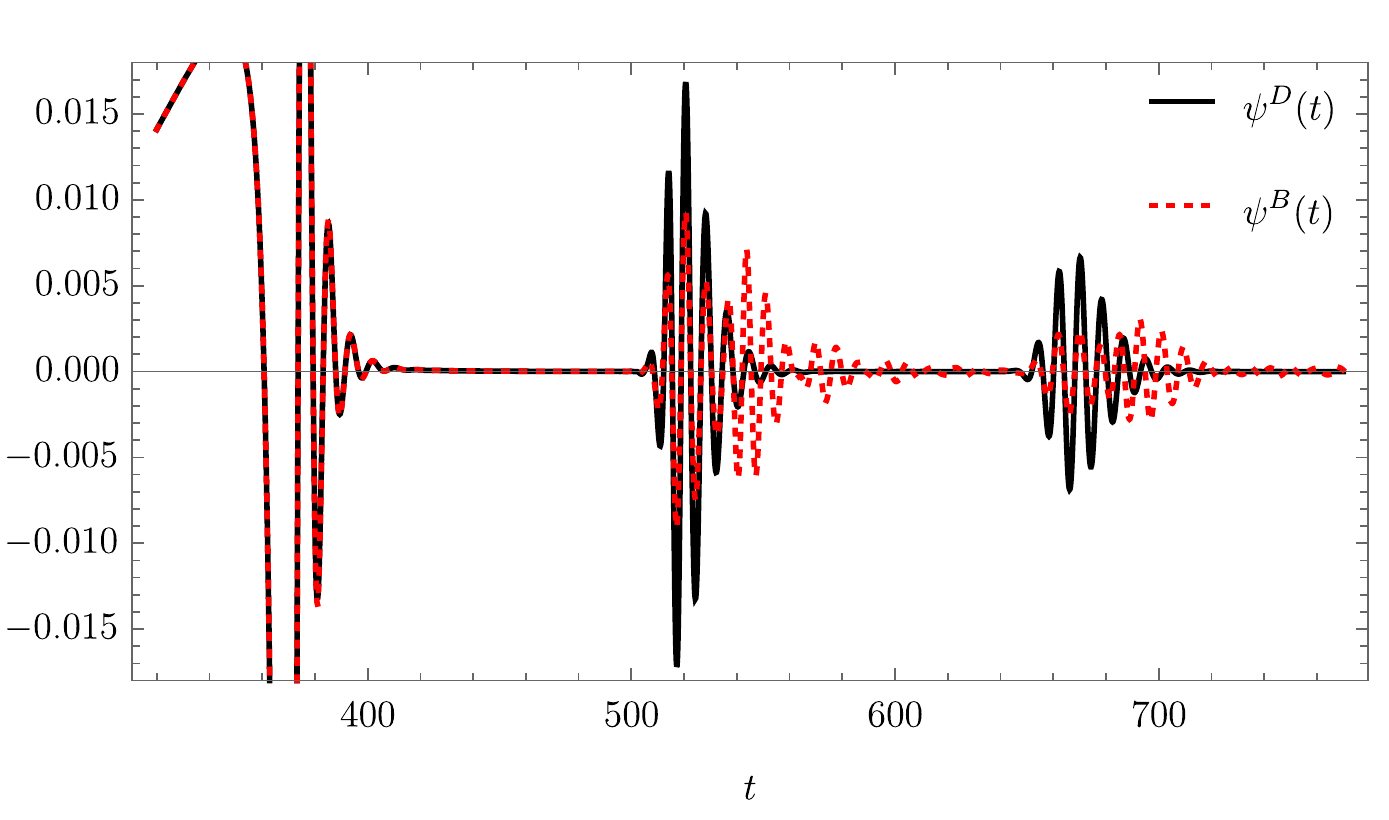}
\caption{Scattering of an incoming Gaussian wave packet centered around $x=x_c=220$ at $t=0$ off  the Schwarzschild potential \eqref{veff} ($s=l=2$), as seen by an observer at $x=150$ (we adopt $2M=1$ units).  The physics is modified relative to Schwarzschild at the near-horizon surface $x=x_\epsilon = -70$.  The black, solid line shows the effects of imposing Dirichlet boundary conditions at  $x=x_\epsilon$ (case {\bf D} in main text). The red, dotted line shows the effects of adding a double square barrier potential near $x=x_\epsilon$ (as described in the text, case {\bf B}). The parameters are given in Eq.\ \eqref{params} and \eqref{scatter_params}. The initial ringdown signal around $t \approx x_c + 150 = 370$ coincides for both cases as it only depends on the Schwarzschild potential near the photon sphere at $r = (3/2)r_s$ (which with our parameters is $x \approx .8$), while the reflecting surface at $x_\epsilon = -70$ is still causally disconnected. The first and second echoes arrive after $\Delta t \sim 2 |x_\epsilon| \approx 140$ and $\Delta t \sim 4 |x_\epsilon|$, respectively. Relative to the Dirichlet case the distortion due to the partially absorbing character of the double square well is visible, as well as the effects of longer-lived  modes trapped between the two barriers.
}
\label{fig:psixt_echo}
\end{figure}

The time evolution of both solutions is shown in Fig.\ \ref{fig:psixt_echo}.
The outgoing, right-moving component of the initial Gaussian wave is discarded.
The ingoing left-moving component of the initial data eventually reaches the photon sphere, where it excites the photon sphere modes \cite{Cardoso:2016rao,Cardoso:2017njb}. This interaction is responsible for the initial ringdown signal, visible until $t\sim 420$. The prompt contribution and photon sphere modes are insensitive (due to causality) to the boundary conditions at  $r = r_\epsilon$. Thus, a classical BH gives rise to exactly the same initial response.

After a time $\sim 2|x_\epsilon| \sim 2 \log\epsilon^{-1}$ following the initial ringdown, the waves have reached  the boundary at $x_\epsilon$ and returned to the observer. The outgoing pulse interacts again with the photon sphere: a fraction transmits out and gives rise to a second pulse of radiation. The remaining fraction is reflected inwards, and the process repeats. A series of echoes is released from the confining region (i.e., between the photon sphere and ``surface'' of the BH $r_{\epsilon}$).  The precise shape and spectral content of each pulse is related to the boundary conditions at $r_{\epsilon}$. Dirichlet conditions, which also correspond to a vanishing flux at the surface, preserve the shape of the reflected wave. On the other hand, the double barrier removes certain frequencies each time the signal interacts with the boundary, leading to characteristically distorted echoes.

\subsection{Quantifying and comparing the filtering\label{filtering_compare_gaussian}}

\begin{figure}[t]
	\begin{tabular}{cc}
		\includegraphics[]{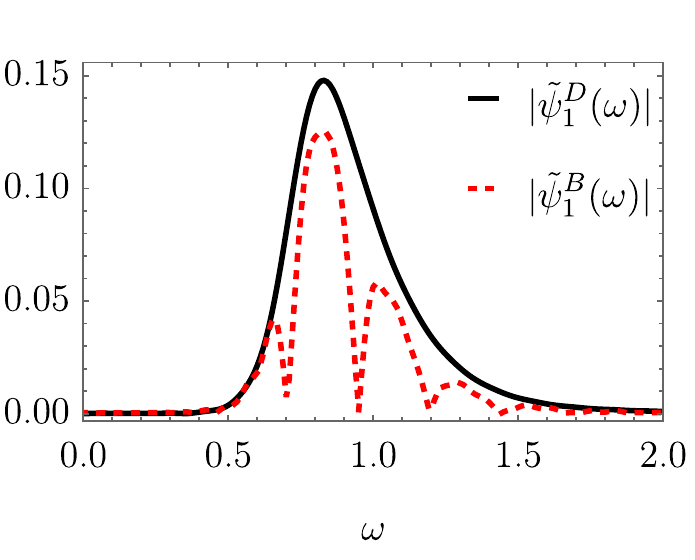} & \includegraphics[]{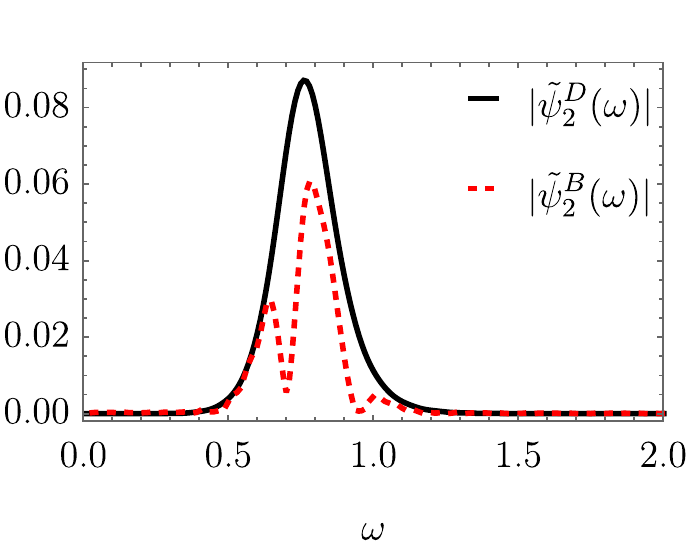}
	\end{tabular}
	\caption{The norm of the Fourier transform \eqref{echo_fourier} of the first (left) and second (right) echo computed numerically as a function of frequency.  The case of Dirichlet boundary conditions is shown in solid black, double barrier in dotted red (we adopt $2M=1$ units). The dips correspond to the filtered frequencies that leaked through the barrier into the BH, and they coincide with the reflection coefficient (see Fig.~\ref{fig:echoes_spectra}).  }
	\label{fig:1stechofourier}
\end{figure}
\begin{figure}[t]
\begin{tabular}{cc}
\includegraphics[]{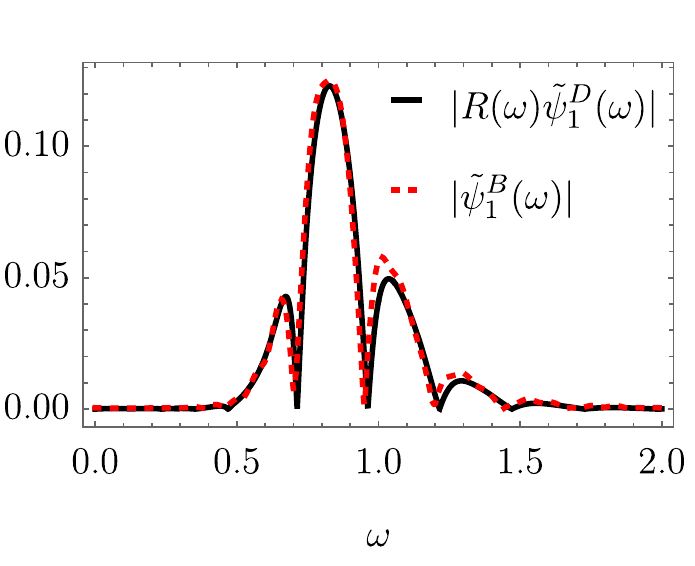}&\includegraphics[]{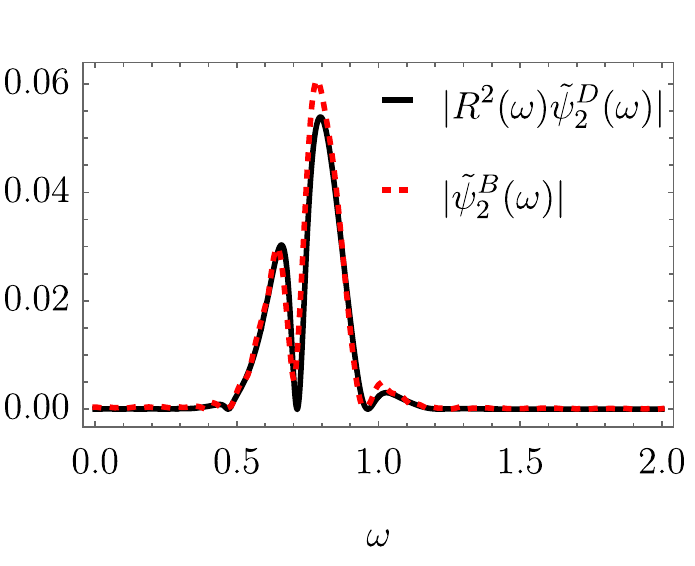}
\end{tabular}
\caption{Comparison of the Fourier transform of the first two echos obtained  either by numerically solving the double barrier boundary condition and obtaining $\tilde{\psi}_{k}^{B}$, or through filtering:
multiplying the Fourier transform of the numerical Dirichlet signal $\tilde{\psi}_{k}^{D}$ by $R^k$, where $R$ is the analytic reflection amplitude for the double square barrier. 
The results agree very well and establish the validity of this approximation.  The latter procedure could be used with any $R$ derived from a theory of the physics near a black hole horizon.  Conversely, if echoes are detected, their Fourier transform can be used to measure $R$ and compare it to the predictions from black hole area quantization. }
\label{fig:echoes_spectra}
\end{figure}
The double barrier  prescription of Section \ref{subsec:doublebarrier} yields the waveform in Fig.\ \ref{fig:psixt_echo}. To implement the procedure described in Section \ref{subsec:bandpass}, we first isolate each echo in the time domain. Suppose that the $k$th echo has support between the times $t^k_1$ and $t^k_2$. The first step is to compute the   Fourier transform of the echoes:
\begin{align}
	\label{echo_fourier}
	\tilde{\psi}_{k}^{D,B}(\omega) = \int_{t^{k}_1}^{t^{k}_2} \psi^{D,B}(t) ~ e^{i \omega t} ~ d t\,.
\end{align}
Note that this procedure assumes the echoes are sufficiently separated from each other. 
The frequency content of the first and second echoes for the Dirichlet and double barrier cases computed with this definition are shown in Fig.\ \ref{fig:1stechofourier}.
We then take the Fourier transform of the $k$th echo of the Dirichlet wave $\tilde{\psi}_{k}^{D}$ and simply multiply by the reflection coefficient of the boundary at $r_\epsilon$, raised to the power $k$ \cite{Mark:2017dnq}. The filtered spectrum is given by
\begin{align}\label{nr_B_spectrum}
	 \tilde{\psi}_{k}^{D'}(\omega) = R^k(\omega) \tilde{\psi}^{D}_{k}(\omega),
\end{align}
where the prime denotes filtering.
The factor $R^k$ takes into account the $k$ times the wave packet bounced off the boundary. This gives us a Fourier transformed echo that should be a good approximation to $\tilde{\psi}_{k}^{B}$.

We use for $R(\omega)$ the analytic reflection coefficient for the double barrier potential with the same parameters \eqref{params} as for the time evolution.
The outcome of this algorithm is shown in Fig.~\ref{fig:echoes_spectra}, where we compare against the original Fourier transform $\tilde{\psi}_{k}^{B}$. The agreement is  good, validating our procedure. 

The real time waveform of any filtered signal is  given by the inverse Fourier transform of the filtered spectrum
\begin{equation}\label{nr_B_waveform}
	\psi'_k(t) = \frac{1}{2\pi} \int_{-\infty}^\infty R^k(\omega) \tilde{\psi}_k(\omega) ~ e^{- i \omega t} ~ d \omega.
\end{equation}
Alternatively, one can work with the Fourier transformed reflectivity and utilize the convolution theorem
\begin{equation}
\psi'_k(t) = (R * \ldots R * \psi_k)(t),
\end{equation}
where the convolution is repeated $k$ times and each convolution is given by $(f * g)(t) = \int_{-\infty}^{\infty} f(\tau) g(t-\tau) d \tau$.

\section{Echoes from black-hole mergers\label{sec:inspiral}}
\subsection{Simulation of extreme-mass-ratio inspirals}

\begin{figure}[t]
\centering
\includegraphics[]{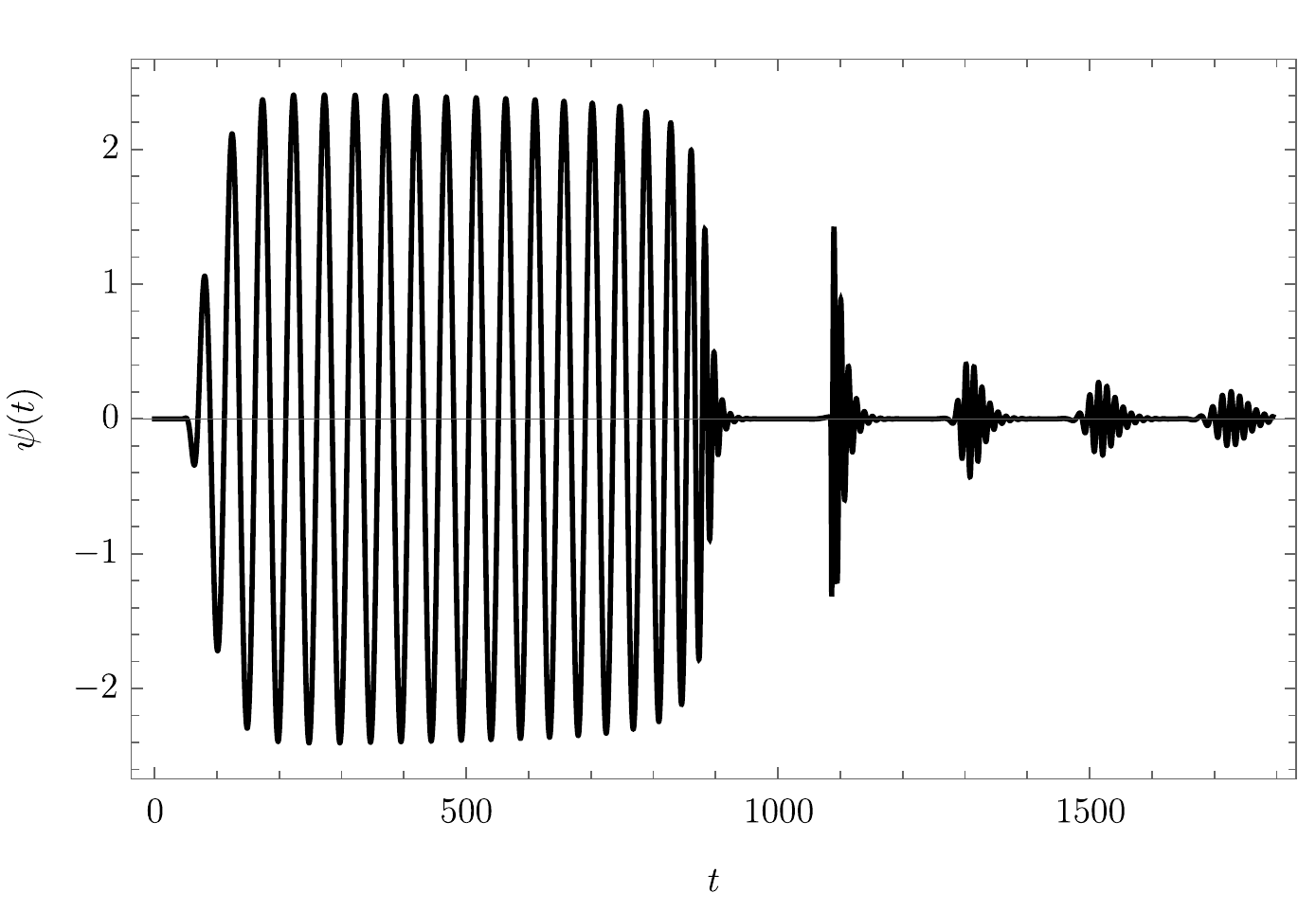}
\caption{Inspiral, merger, and ringdown waveform of the numerical relativity simulation of an inspiral with mass ratio $M_1/M_2 = 10^3$, with Dirichlet boundary conditions near the horizon, and the first four of the resulting echoes.
}
\label{fig:numeric_gr}
\end{figure}
The previous discussion dealt with the scattering of radiation pulses off BH spacetimes. A more relevant process to  astrophysical observations is the inspiral and merger of two BHs, a process which releases an enormous amount of energy through gravitational waves, which excite the modes of the final BH to a considerable amplitude~\cite{Berti:2009kk,Berti:2016lat,LIGOScientific:2018mvr,Barack:2018yly}. The simulation of equal or nearly-equal mass BHs is challenging and requires the full use of numerical relativity techniques. However, one can more easily simulate the inspiral and merger of extreme-mass ratio inspirals, a process which is itself of great interest for future detectors such as LISA~\cite{Audley:2017drz,Barack:2018yly}. The inspiral and merger of a small star or BH with a massive one can be simulated with perturbation techniques, which include dissipation of energy and which drive the inspiral~\cite{Khanna:2016yow,Price:2017cjr,Tsang:2018uie}. 
The dominant, quadrupolar mode generated when a BH with mass $M_2 = 10^{-3}$ inspirals onto a large BH with mass $M_1=1$ is shown in Fig.~\ref{fig:numeric_gr}. Here, reflecting Dirichlet boundary conditions were set at $x_\epsilon=-100$, which leads to the emission of a wave train of echoes.

We  now study the effect of non-trivial boundary conditions on those echoes.
In other words, we compute the change in the waveform when the Dirichlet boundary condition at $x_\epsilon = -100$ is replaced by a surface with arbitrary reflection coefficient $R(\omega)$. We can use the procedure outlined in Section \ref{subsec:bandpass}, which has been tested in the previous section. This allows us to compute both the spectrum of every echo in the series, as well as the distorted waveforms that could be detected in future gravitational wave experiments.

\begin{figure}[t]
	\begin{tabular}{cc}
		\includegraphics[width=0.45\textwidth]{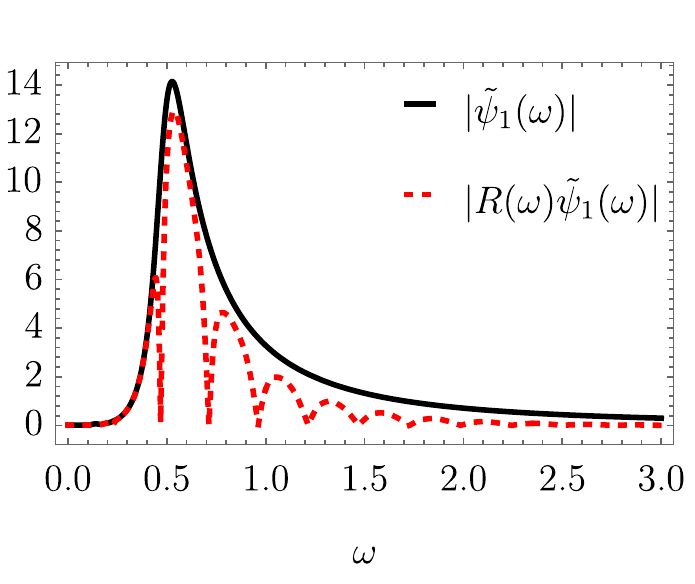}&\includegraphics[width=0.45\textwidth]{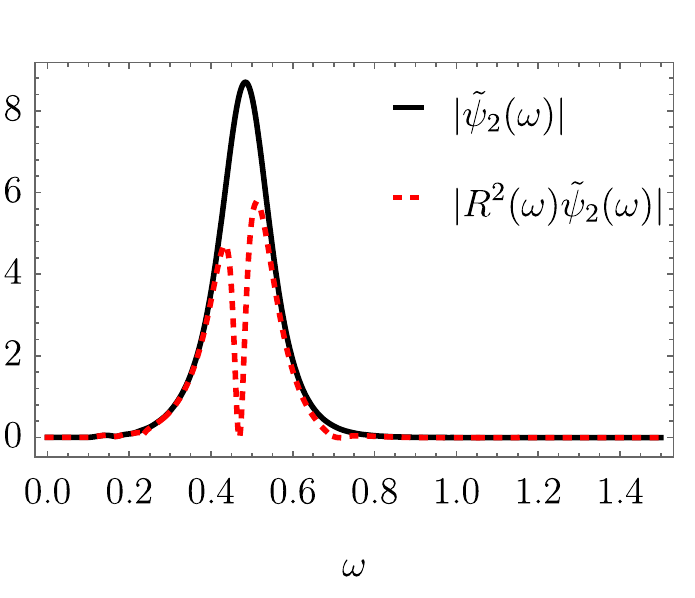}
	\end{tabular}
	\caption{The spectrum \eqref{nr_B_spectrum} for the first (left) and second (right) echo as a function of frequency, Dirichlet boundary conditions in solid black, filtered signal in dotted red. The double barrier reflection coefficient with parameters $d=12, L=1, V_0=1$ were used for the filtering.
	}
	\label{fig:gr_b_echoes}
\end{figure}

\begin{figure}[t]
	\begin{tabular}{cc}
		\includegraphics[width=0.45\textwidth]{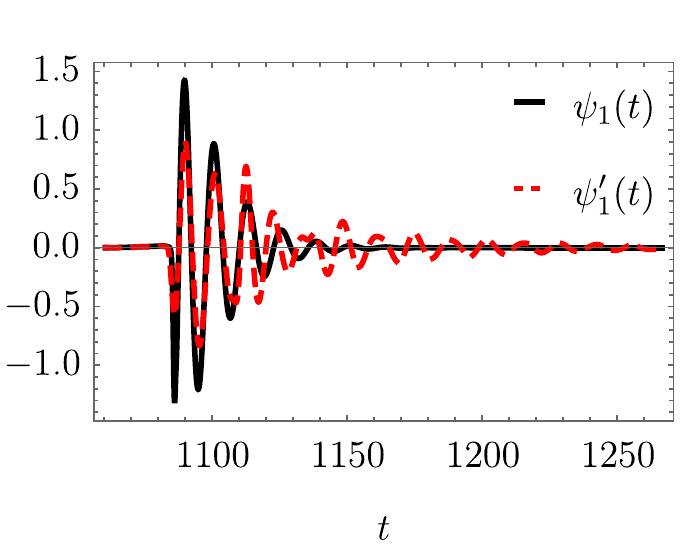}&\includegraphics[width=0.45\textwidth]{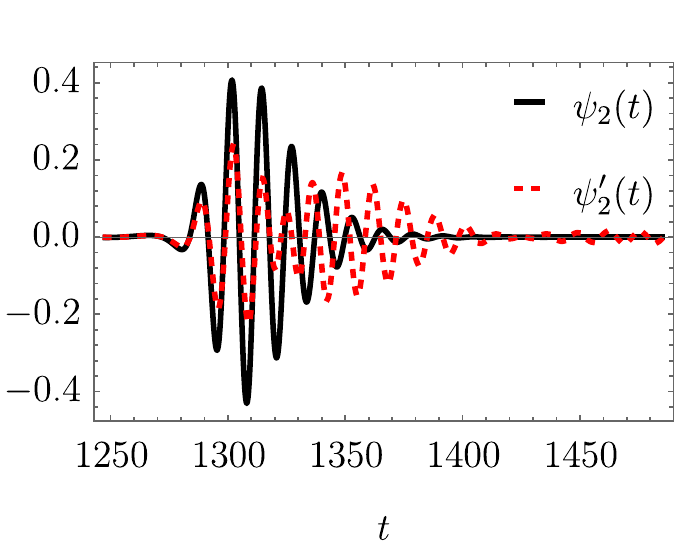}
	\end{tabular}
	\caption{The real time waveform \eqref{nr_B_waveform} of the same first two echoes displayed in Fig.~\ref{fig:gr_b_echoes}: filtered by the double barrier boundary conditions at $r_\epsilon$ (dotted red) together with the unfiltered (Dirichlet boundary conditions at $r_\epsilon$) signal (solid black). The partial absorption at $r_\epsilon$ introduces distortion and longer-lived modes (see Fig.~\ref{fig:psixt_echo}).
	}
	\label{fig:gr_b_waveforms}
\end{figure}

Fig.\ \ref{fig:numeric_gr} shows that the echoes are  sufficiently separated to identify the start and end times of the $k$th echo, which we again denote by $t_1^k$ and $t_2^k$.  We repeat the analysis for the reflection coefficient of the double barrier from the previous section.
The frequency content of each echo is again given by their Fourier transforms \eqref{echo_fourier} with appropriate start and end times.
The spectra of the first two echoes are shown in Fig.\ \ref{fig:gr_b_echoes}. The real time waveform of the echo and the filtered signal according to \eqref{nr_B_waveform} are shown in Fig.\ \ref{fig:gr_b_waveforms}.

\section{Discussion}

So far we have  addressed the case where the distance between the photon sphere and the reflective boundary at $x_\epsilon$ is large enough for individual, well separated echoes to emerge. If the trapped region is smaller, echoes start to overlap and we are dealing with echo interference. This case is tractable with an additional step. Instead of filtering the waveform scattered back to infinity, one has to work with the incoming radiation and consider the filtering through the effective Schwarzschild potential separately. Illustrative examples are presented in \cite{Mark:2017dnq}.

The leading order effect is the triggering of a series of postmerger echoes. In our analysis -- which deals only with non-spinning backgrounds -- each echo is damped compared to the previous, due to the gradual energy loss to infinity and into the horizon. There is one important qualitatively new aspect in rotating backgrounds: the presence of ergoregions will lead to exponentially {\it growing} echoes
in the initial stages where linear analysis is still valid~\cite{Vicente:2018mxl,Brito:2015oca,Bueno:2017hyj,Maggio:2018ivz,Barausse:2018vdb,Wang:2018gin}. 

When $\epsilon \ll 1$ the time between echoes is large ($\approx 4M \ln \epsilon^{-1}$).  In other words, the ``cavity'' between the photon sphere and the modified horizon has a small natural frequency.  It is conceivable that this cavity could be excited during the period of the inspiral phase when the orbital frequency resonates with it.  Such effects  are outside the scope of this work.

In the limit where the line width is very small, the horizon will act as a nearly perfect reflector. The emergence of echoes would still signal exotic new physics, however, in that case the nontrivial absorption near the horizon could be challenging to detect.

\section{Conclusions}

According to the proposal of Bekenstein and Mukhanov, the area of black horizons is quantized in units of the Planck area.
The frequency gaps between these levels are of order the inverse black hole radius, and their width is  assumed to be small compared to the gaps.  
Rather than being a perfect absorber, the horizon acts as a filter with a set of absorption lines.  
The main outcome of our analysis is the nontrivial postmerger signal
produced when two compact objects inspiral and merge to form a single BH. Our analysis predicts that the classical GR signal
would be followed by a sequence of ``echoes'' following the main burst (Fig.~\ref{fig:numeric_gr}), deformed in a specific way due to the selective absorption at the horizon.

\acknowledgments
V. C. acknowledges financial support provided under the European Union's H2020 ERC 
Consolidator Grant ``Matter and strong-field gravity: New frontiers in Einstein's 
theory'' grant agreement no.\ MaGRaTh--646597.
This project has received funding from the European Union's Horizon 2020 research and innovation program under the Marie Sklodowska-Curie grant agreement No 690904.
We acknowledge financial support provided by FCT/Portugal through grant PTDC/MAT-APL/30043/2017.
We acknowledge the SDSC Comet and TACC Stampede2 clusters through NSF-XSEDE Award Nos.\ PHY-090003.
The authors would like to acknowledge networking support by the GWverse COST Action 
CA16104, ``Black holes, gravitational waves and fundamental physics''.
The authors thankfully acknowledge the computer resources, technical expertise and assistance provided by CENTRA/IST. Computations were performed at the cluster ``Baltasar-Sete-S\'ois'' and supported by the H2020 ERC Consolidator Grant ``Matter and strong field gravity: New frontiers in Einstein's theory'' grant agreement no.\ MaGRaTh-646597.
The work of MK is supported by the NSF through grants PHY-1214302 and PHY1820814.  The authors thank the ICTP, where this work was initiated, for its hospitality.

\bibliographystyle{JHEP}
\bibliography{bibliography}

\end{document}